\documentstyle[psfig,11pt]{article}
\begin{document}
\vspace{1.0cm}
{\Large \bf BeppoSAX Observations of the Radio Galaxy Centaurus A}

\vspace{1.0cm}

P. Grandi$^1$, C. M. Urry$^2$, L. Maraschi$^{3}$, M. Guainazzi$^4$, 
E. Massaro$^{5}$, G. Matt$^{6}$, L. Bassani$^7$, A. Cimatti$^8$, 
P. Giommi$^9$, M. Dadina$^{9}$, G. C. Perola$^{6}$, L. Piro$^1$, 
A. Santangelo$^{10}$.

\vspace{0.5cm}
$^1${ IAS/CNR, Area di Ricerca Tor Vergata,
Via Fosso del Cavaliere, I-00133 Roma, Italy }\\
$^2${ STScI, 3700 San Martin Drive, Baltimore, MD 21218, USA}\\
$^3${ Osservatorio Astronomico di Brera, Via Brera 28, I-20121 Milano, Italy}\\
$^4${ SSD/ESA, ESTEC, Postbus 299, 2200 AG Noordwijk, The Netherlands}\\
$^{5}${ Istituto Astronomico di Roma, Via Lancisi 29 Roma, Italy}\\
$^{6}${ Universit\`a degli Studi ``Roma 3'', Via della Vasca 
Navale 84, I-00146 Roma, Italy}\\
$^7${ ITESRE/CNR, Via P. Gobetti 101, I-40129, Bologna, Italy}\\
$^8${ Osservatorio Astronomico di Arcetri, Largo E. Fermi 5, I-50125, 
Firenze, Italy}\\
$^9${ BeppoSAX SDC, c/o Nuova Telespazio, Via 
Corcolle 19, I-90146, Roma, Italy}\\
$^{10}${IFCAI/CNR, Via U. La Malfa 153, I-90146, Palermo, Italy}\\
\vspace{0.3cm}

\section*{ABSTRACT}

We present preliminary results from two observations of
the radio galaxy Centaurus~A performed by the BeppoSAX satellite on
1997 February 20-21 and on 1998 January 6-7.
In the second pointing the source was brighter by a factor 1.3.
We did not detect any spectral variation of the nuclear continuum 
in spite of the long-term flux change between the two observations.
At both epochs, the nuclear point-like emission was well fitted 
with a strongly absorbed ($N_H\sim10^{23}$ cm$^{-2}$)
power law with an exponential cutoff at high energies
(E$_{cut}> 200$~keV). 
We also observed a significant flux variation of the iron line between
the two observations. The flux of the line and of the continuum 
changed in the opposite sense. The line is more intense at the first epoch,
when the nuclear source was at the lower intensity level. 
The implied delay between the continuum and line variations strongly suggests
that the cold material responsible for the iron line production 
is not located very near to the primary X-ray source.
There is also evidence that the line profile changed between the two
epochs, being broader and slightly blueshifted when the source was fainter.
It is possible that the emission feature is a blend of cold and 
ionized iron lines produced in separate regions surrounding the 
nuclear source.

\section{INTRODUCTION}
\vspace{-5mm}
The radio galaxy Centaurus~A (Cen~A) is a well-known radio-loud AGN
which has been studied extensively over the whole X-ray and $\gamma$-ray range.
Its proximity (z=0.008) means its X-ray emission has been spatially resolved,
into at least five different regions:
the nucleus, the jet, a middle NE radio lobe between 
20$^\prime$-25$^\prime$ from the nucleus,
a faint diffuse emission within 6$^\prime$ probably from hot interstellar medium
and two ridges of thermal emission along each edge of the dust line
(Feigelson et al. 1981, Turner et al. 1997).
Above 3~keV the spectral emission is essentially dominated
by the nuclear point-like source, while at soft energies the extended
thermal components become important because of the large extinction
suffered by the nucleus.
The X-ray spectrum between 3-10~keV is generally modeled by a 
power law heavily absorbed by dense and probably stratified matter 
and by a fluorescence iron line (Morini et al. 1989, Miyazaki et al 1996, 
Turner et al. 1997, Sugizaki et al. 1998). 
Hard X-ray and soft $\gamma$-ray observations performed with the GRO satellite
show a steepening of the power law at energies larger than 120~keV
(Kinzer et al. 1995, Steinle et al. 1997).

Here we present preliminary results of two observations of Cen~A
performed with the BeppoSAX satellite. We concentrate 
on the nuclear X-ray emission, giving information on the continuum 
spectral shape and on the iron line feature.

\section{RESULTS}
\vspace{-5mm}
The Narrow Field Instruments (LECS: 0.1-10~keV; MECS: 1-10~keV;
HPGSPC: 4-120~keV; PDS: 15-300~keV) of BeppoSAX (Parmar et al. 1997, 
Boella et al. 1997, Manzo et al. 1997, Frontera et al. 1997) observed 
Cen~A twice on 1997 February 20-12 and 1998 January 6-7 
for 35 and 75 ksec, respectively.
The average source flux increased between the two observations, being 
brighter by about a factor 1.3 in 1998.
The brightness level was however relatively low in both 
observations compared to the historic light curve.
The observed flux, $F_{\rm 3-12~keV}\sim 0.2-0.3\times10^{-9}$ photons
cm$^{-2}$ sec$^{-1}$, is within the range observed in the last 10 years
and lower at least by a factor $\sim 5$ than the outburst in 1974-75
(Turner et al. 1997). 

Here we present the results of the preliminary analysis of the MECS and 
PDS instruments, covering the 1.5-150 keV energy band.
The MECS spectra were accumulated with 8$^\prime$ extraction radii.
Because of a failure in the MECS unit 1 on 1997 May 6 
only data from two units were available in the 1998 observation.
The extraction region of the MECS was large enough to include extended
components, in particular the X-ray emission from the galactic 
ridges and from the jet. The diffuse thermal ($kT=0.9$~keV) emission produced 
by interstellar medium is very soft and therefore negligible in the 
1.5-150~keV range. 
A spatially resolved spectroscopic study based on ROSAT observations (Turner et al. 1997) 
has shown that the galactic ridge emission can be well fitted with 
a Raymond-Smith model ($kT=5$~keV, metal abundances 0.4 
times solar) modified by Galactic absorption 
($N_H=7\times10^{20}$ cm$^{-2}$), and the jet emission with an 
absorbed power law (N$_H=9.7 \times10^{20}$ cm$^{-2}$, $\Gamma=2.29$).
We used those parameters to model the extended emission components and 
used the observed luminosity corrected by absorption 
($L_{\rm jet}^{\rm 0.1-2~keV}=2\times10^{39}$ erg sec$^{-1}$; 
$L_{\rm gal. ridges}^{\rm 0.1-2~keV}=2.4\times10^{39}$ erg sec$^{-1}$) to 
set the normalization constants.
\begin{table*}[t]
\begin{center}
\footnotesize
\caption[] {\it MECS+PDS fit results. The contributions from the galactic ridges 
and the jet are parameterized by a Raymond-Smith model ($kT=5$~keV and 
0.4 $\times$ solar abundances, Galactic absorption) and an absorbed power law 
($\Gamma=2.29$, N$_H=9\times10^{20}$ cm$^{-2}$),
respectively (see text).
The quoted errors are $90\%$ confidence for a single parameter.}
\begin{tabular}{lccccccccc}
\noalign {\hrule}
&&&&&&&&&\\
Model & A$^a$ &$N_{H}$ & $\Gamma$ & E$_{\it cutoff}$
& R& E$_{\it Fe}$ & $\sigma$ & F$_{\it Fe}^b$ & $\chi^2(dof)$\\
&& $10^{22}$ cm$^{-2}$& & (keV) & & keV & keV &&\\
&&&&&&&&&\\\hline
&&&&&&&&&\\
\multicolumn{10}{c}{1997 February 20-21}\\
&&&&&&&&&\\
PL  &8.8$\pm0.3$&9.9$\pm0.2$ & 1.77$^{+0.02}_{-0.01}$& ... &...
 &6.49$^{+0.07}_{-0.08}$& 0.3$\pm0.1$& 5.1$^{+1.1}_{-0.9}$&124(89)\\
CPL & 7.7$\pm0.4$ & 9.5$\pm0.2$& 1.68$\pm0.03$&
215$^{+107}_{-55}$& 0.0 
& 6.47$^{+0.06}_{-0.07}$ & 0.3$\pm0.1$ & 
4.8$^{+1.1}_{-0.8}$& 97(88) \\
CPL+ REF  & 8.5$^{+0.9}_{-0.6}$ & 9.8$^{+0.2}_{-0.3}$ & 1.76$^{+0.04}_{-0.06}$
& 286$^{+257}_{-94}$& 0.21$^{+0.05}_{-0.14}$&6.48$\pm0.07$  & 0.3$\pm0.1$
& 5.2$^{+1.2}_{-1.0}$&90(87)\\
&&&&&&&&&\\\hline
&&&&&&&&&\\
\multicolumn{10}{c}{1998 January 6-7}\\
&&&&&&&&&\\
PL  &11.9$\pm0.2$&9.6$^{+0.1}_{-0.2}$ & 1.78$\pm0.01$&... &...  &
6.39$^{+0.07}_{-0.08}$& 0.07$^{+0.22}_{-0.07}$ & 2.5$^{+1.1}_{-0.7}$ & 147(89)\\
CPL &10.5$\pm0.4$&9.2$\pm0.2$& 1.70$\pm0.02$&250$^{+81}_{-50}$& 0.0 & 
6.38$^{+0.07}_{-0.08}$ & 
0.06$^{+0.23}_{-0.06}$ &2.5$^{+0.8}_{-0.7}$ &98(88) \\
CPL+REF &11.1$^{+0.9}_{-0.7}$ & 9.2$^{+0.3}_{-0.1}$ & 1.73$^{+0.05}_{-0.04}$& 
297$^{+156}_{-79}$
& 0.08$^{+0.09}_{-0.08}$& 6.38$\pm0.08$  & 0.08$^{+0.23}_{-0.08}$
& 2.6$\pm0.8$  & 95(87)\\
&&&&&&&&\\
\hline
\multicolumn{10}{l}{$^a$ Photons $\times10^{-2}$ cm$^{-2}$ sec$^{-1}$ 
keV$^{-1}$}\\
\multicolumn{10}{l}{$^b$ Photons $\times10^{-4}$ cm$^{-2}$ sec$^{-1}$}\\

\end{tabular}
\end{center}
\end{table*}

With the contributions of the galactic ridges and the jet fixed, 
we studied the spectral shape of the nuclear point-like component.
In Table~1 are listed the spectral fit results from both observations.
An absorbed power law (PL) was not an acceptable model (see Table~1 and 
Fig.~1) for the continuum in both the observations.
More reasonable fits were obtained with a cutoff power law (CPL).
The CPL fit to the 1997 data improved significantly ($P_{F_{\it test}}=98.9\%$)
if a reflection component was added to the model (R in Table~1 is
the ratio between the reflection and power law normalizations).
\begin{figure}[b]
\centerline{\psfig{figure=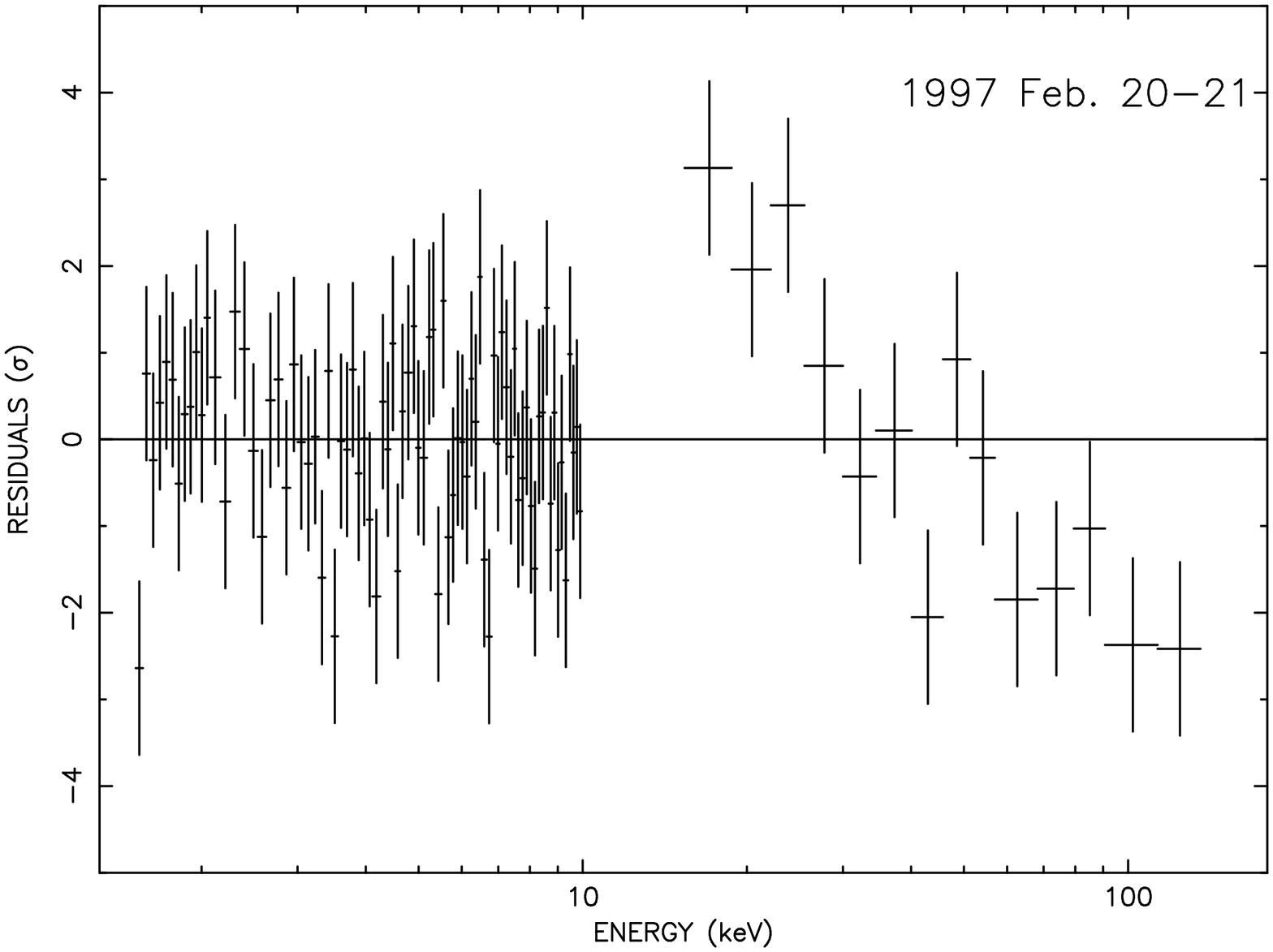,height=5.5cm,width=7.5cm,angle=-90}
\psfig{figure=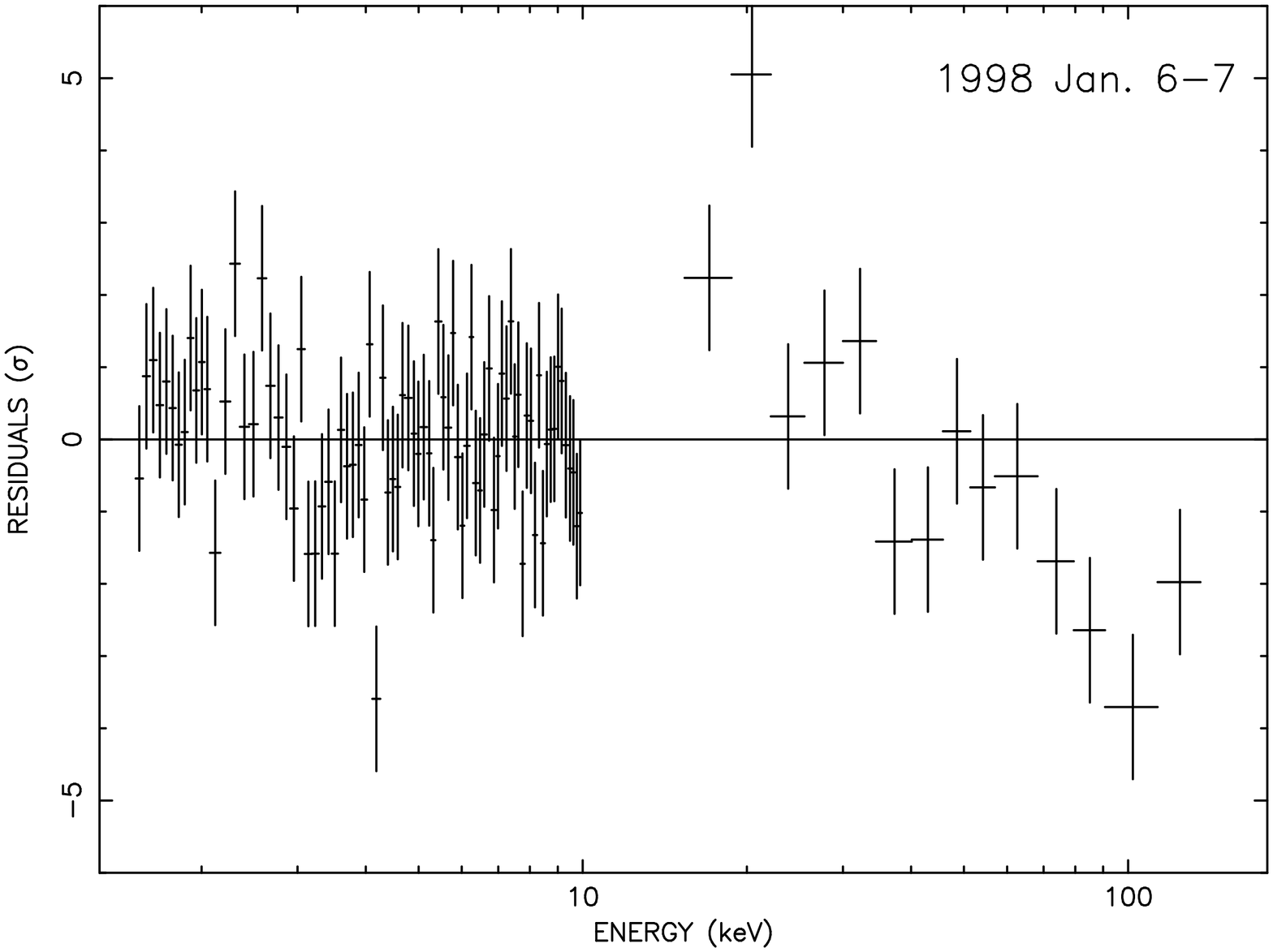,height=5.5cm,width=7.5cm,angle=-90}}
\caption{Residuals to the 1997 (left panel) and 1998 (right panel) 
spectra of the Cen~A nucleus after fitting with
a simple power law model. A Gaussian profile was used to model the iron line feature.}
\end{figure}
In the 1998 data, characterized by a larger nuclear flux,
the inclusion of a reflection component gave a less significant improvement 
($P_{F_{\it test}}=94\%$) and R was consistent with zero.
It is possible that reprocessed X-ray radiation
was present in the first observation and not in the second, as
can occur if the X-ray reflecting region is far from the nuclear source.
However, since the amount of reflection is strongly dependent on the
relative flux normalization adopted for the MECS and PDS instruments, 
which have intercalibration uncertainties of $\sim 3\%$, and since
the amount of reflection ($R=0.2$) is quite small, we can not exclude that
the reflection component may be produced by miscalibration effects.
\noindent
Comparing the results in Table~1, we can conclude that 
the spectral shape of the primary X-ray radiation did not change 
significantly with the intensity.

A Gaussian profile fits the iron line in both observations, but
in contrast to the primary X-ray continuum spectrum, the emission feature 
changed shape.
Figure~2 (left panel) shows the ratio between the 1997 and 1998 spectra.
It is evident that the ratio is constant across the entire 
3-150~keV range apart from the iron line region.
The iron line was indeed clearly more intense when the source was fainter.
As shown in Table~1, the total number of photons in 
the line was about twice as large in the first observation, when 
the nuclear flux was $\sim 25\%$ lower.
This result implies that the region responsible for the 
iron line responds with a significant delay to the 
continuum variations and therefore 
the reprocessing material cannot be located very near to the primary 
X-ray source.
The data also indicate that the shape of the 
Gaussian profile changed. The iron line appears broader and 
slightly blueshifted in the first observation relative to the second
(Fig. 2, right panel).
Note that, while the iron line is well resolved in the 1997 
observation ($\sigma=0.3\pm0.1$ keV), its width is consistent with zero 
in 1998.

One possibility is that the 1997 feature was a blend of 
cold and ionized lines.
As often observed in Seyfert 2 galaxies, the 6.4 keV line might be 
produced by a molecular dusty torus and the ionized one by warm 
scattering material out of the line of sight. 
Alternatively, the reprocessing material might be stratified and 
roughly spherical with the ionized region closer to the nuclear source.
The presence of the reflection component when the iron line flux was larger,
even if marginal, 
seems however to favour a toroidal 
geometry for the cold absorber.

\begin{figure}[th]
\centerline{\psfig{figure=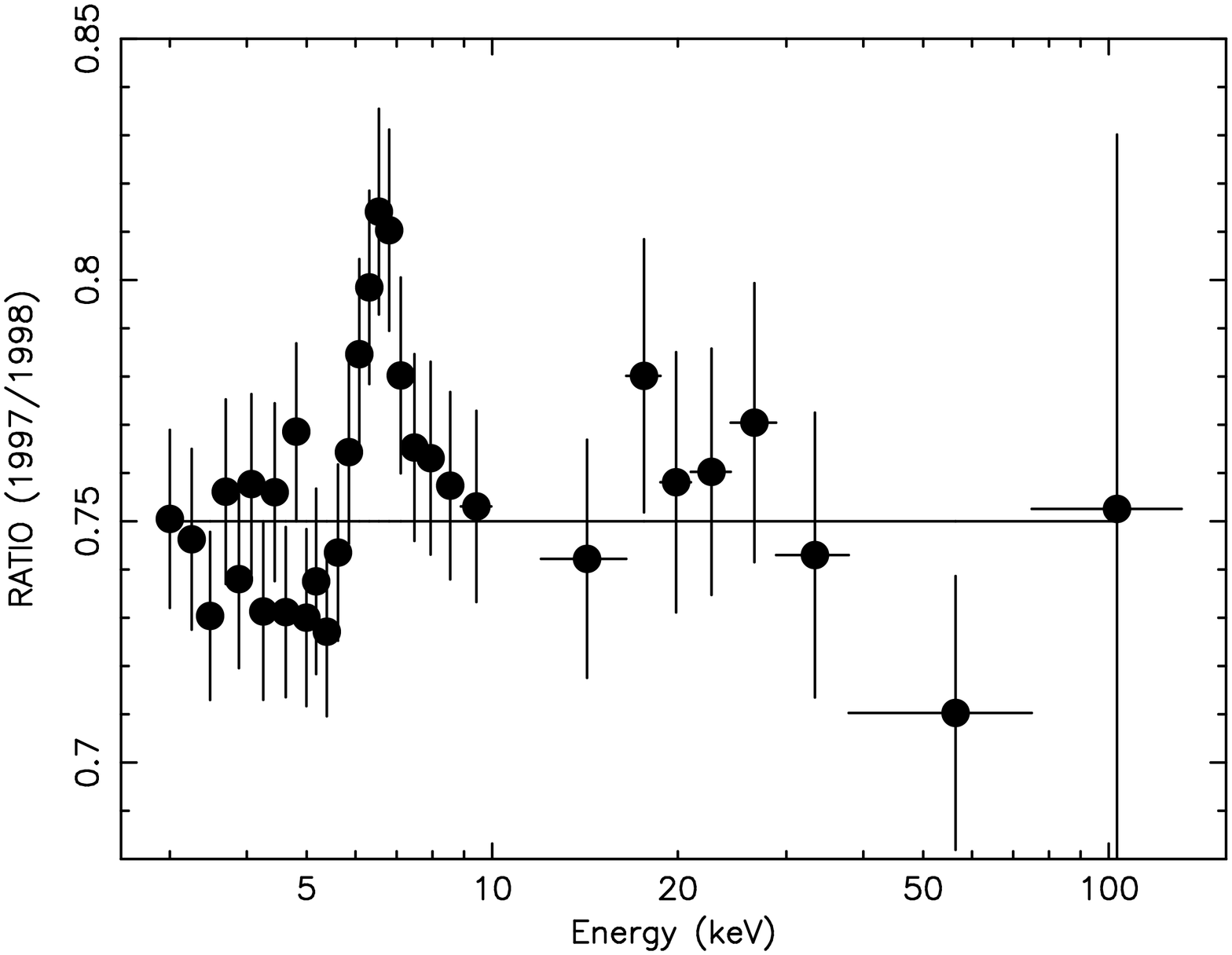,height=5.5cm,width=7.5cm,angle=-90}
\psfig{figure=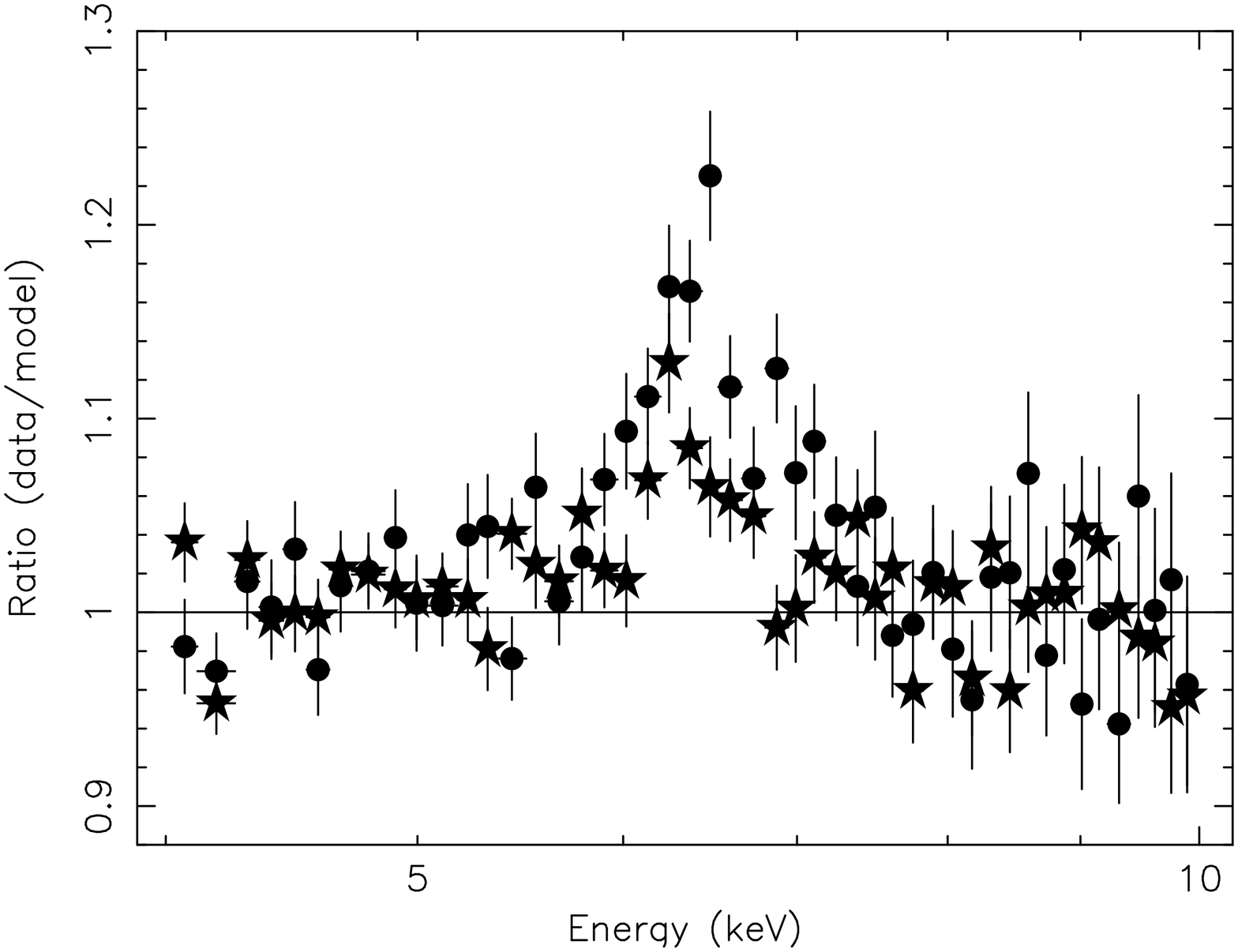,height=5.5cm,width=7.5cm,angle=-90}}
\caption{({\it Left Panel}) -- The 1997 data divided by the 1998 
data. MECS data of unit 1, collected during the first observation, 
are not included. The ratio is constant across the 3-150~keV band 
but not in the 5-7~keV iron-line range.
({\it Right Panel}) -- The MECS spectra in the iron line region, after
division by a continuum model (CP+REF) fitted to the 1.5-5.0~keV and 
7.5-150~keV bands.
The filled points refer to 1997 data, the stars to the 1998 observation.}
\end{figure}

\section{CONCLUSION}
\vspace{-5mm}
The long term variability of Cen~A observed with BeppoSAX offered a 
unique opportunity to study the spectral variations of the nuclear emission 
in the medium and hard X-ray (3-150~keV) bands as a function of the 
intensity.
We observed that: (1) Independent of the brightness of the 
nuclear component, the spectral shape of point-like nuclear
source did not change significantly.
In both observations, the spectrum was a power law 
with an exponential cutoff at energies $> 200$~keV,
heavily absorbed by a large invariable column 
density $N_H\sim10^{23}$ cm$^{\-2}$.
An additional hard (reflection) component was present when
the nuclear source was fainter. Since the amount of reflection was small
and the MECS and PDS intercalibration uncertainties are of the order of
$\sim 3\%$, this result should be considered with caution.
(2) The strength of the iron line flux is not directly correlated with
the intensity of the nuclear source.
We conclude that material responsible for reprocessing 
the primary X-ray radiation is not very near to  the nuclear source.
In addition, the change of line profile 
(intrinsic width and energy peak) with nuclear flux suggests
the feature is a blend of cold and ionized line, probably produced 
by separate regions with different opacities. 

\section{REFERENCES}
\vspace{-5mm}
\begin{itemize}
\setlength{\itemindent}{-8mm}
\setlength{\itemsep}{-1mm}

\item[]
Boella G., et al.,  A$\&$AS,122, 327 (1997)
\item[] 
Feigelson, E. D., et al., ApJ, 251, 32 (1981)
\item[]
Frontera et al., A$\&$AS, 122, 357 (1977)
\item[] 
Kinzer, R. L., et al., ApJ, 449, 118 (1995)
\item[]
Manzo G., et al. A$\&$AS, 122, 341 (1977)
\item[]
Miyazaki, S., et al., PASJ, 48, 801 (1996)
\item[]
Morini, M., F. Anselmo , D. Molteni, ApJ, 347, 750 (1989)
\item[]
Parmar A., et al., A$\&$AS, 122, 309 (1977)
\item[]
Steinle, H., et al., A$\&$A, 330, 97 (1998)
\item[]
Sugizaki, M., et al., PASJ, 49, 59 (1997)
\item[]
Turner, T. J., I. M. George, R. F. Mushotzky, K. Nandra, ApJ 475, 118 (1997)

\end{itemize}
\end{document}